\documentclass[thmsa,a4paper,onecolumn,12pt,final,notitlepage,oneside]{article}
\usepackage{sw20lart}

\input{tcilatex}
\begin{document}

\title{On The Relevance of Fair Sampling Assumption in The Recent Bell Photonic
Experiments}
\author{A. Shafiee\thanks{%
E-mail: shafiee@sharif.edu} $^{\text{(1)}}$ \ and \ M. Golshani$^{\text{(2,3)%
}}\bigskip $\quad \\
%EndAName
{\small \ \ }$\stackrel{1)}{}$ {\small Department of Chemistry, Sharif
University of Technology,}\\
{\small \ P.O.Box 11365-9516, Tehran, Iran}\\
{\small \ }$\stackrel{2)}{}$ {\small Department of Physics, Sharif
University of Technology,}\\
{\small \ P.O.Box 11365-9161, Tehran, Iran.}\\
{\small \ }$\stackrel{3)}{}$ {\small Institutes for Studies in Theoretical
Physics \& Mathematics,}\\
{\small \ P.O.Box 19395-5531, Tehran, Iran.}}
\maketitle

\begin{abstract}
In the experimental verification of Bell's inequalities in real photonic
experiments, it is generally believed that the so-called fair sampling
assumption (which means that a small fraction of results provide a fair
statistical sample) has an unavoidable role. Here, we want to show that the
interpretation of these experiments could be feasible, if some different
alternative assumptions other than the fair sampling were used. For this
purpose, we derive an efficient Bell-type inequality which is a CHSH-type
inequality in real experiments. Quantum mechanics violates our proposed
inequality, independent of the detection-efficiency problems.
\end{abstract}

\section{Introduction}

In his celebrated 1964 paper, John Bell considered a system consisting of
two spin-$\frac{1}{2}$ particles in a singlet state [1]. He showed that the
correlation between the results of two experiments done on such spatially
separated particles cannot be reproduced by a local hidden-variable theory.
Experiments done since 1972 indicate that the spin correlations of two
particles in a singlet state violate Bell's inequality, as quantum mechanics
requires. These experiments have been usually done for photons (see ref. [2]
and the references therein), and recently they were performed for massive
particles [3]. But, there has been two general loopholes in the standard
interpretation of these experiments which are not yet taken care of
simultaneously in a single experiment. They make the present interpretations
inconclusive. These are known as detection loophole and locality loophole.

The detection loophole [4] refers to the fact that in Bell-type experiments,
due to the low efficiency of detectors and collimators, a large number of
photons may be undetected, and the resulting correlation is obtained on the
basis of detected photons. Consequently, it is always possible to construct
a local hidden variable model which can reproduce the experimental results
[4, 5, 6]. So, the interpretation of these experiments is only feasible if
one makes the fair sampling (FS) assumption [7]. P. Grangier describes the
detection-efficiency loophole as ``Achill's heel of experimental tests of
Bell's inequalities'' [8]. In the experiments done with the massive
particles, this problem is solved, because the detection of these particles
could be done efficiently. Yet, the second loophole, i.e., the locality
loophole (according to which there exist the possibility of (sub)luminal
communication between two spatially separated particles) is still there.
(see, e.g., ref. [9].)

In this paper, we consider the problem of fair sampling in the experiments
done with photons. (A complete review of these experiments can be found in
ref. [2].) There have been considerable discussions in the literature on
this subject. But, the main issue in all of them is either to obtain a new
limit for the detector inefficiency in CH [10] and/or CHSH [7] inequalities
(see, e.g., ref. [11]) or to include the detector inefficiency directly into
the original Bell inequality (see, e.g., ref. [12]). Recently, the authors
have also proposed a new CH-type inequality which can be violated by some
quantum mechanical predictions independent of the efficiency factors [13,
14].

What we want to argue here is that, contrary to what is assumed so far,
detection loophole could be neglected in the interpretations of the recent
photonic experiments, since there are other independent assumptions which
are reasonable by themselves and could be used. To show this, we use, in
part 2 of this paper, an \textit{efficient} type of CHSH inequality which
has been tested in recent photonic experiments and is shown to have been
violated. We argue that for observing the violation of this inequality in
real experiments, there are at least three possible solutions (other than
the FS assumption) which have different physical basis and can be used
independently for deriving the inequality. In section 3, we show that
quantum mechanical predictions violate this inequality for real experiments, 
\textit{independent} of the efficiency of detectors and collimators. Thus,
we find \textit{another }way for the justification of the entanglement
criteria in the microphysical Bell states.

\section{An Alternative Bell-type Inequality}

FS assumption means that unrecorded data \textit{do not} have a weighty role
in calculating the polarization correlations of the two entangled photons.
This is the most common view about the FS assumption. That is what P. Pearle
described as Data Rejection Hypothesis in his 1970 paper [4]: ``Suppose that
each particle has three responses to a spin-measuring apparatus instead of
two....Then, instead of four possible experimental outcomes of the
measurement of the spins of two particles, there are nine possible outcomes.
In one of these outcomes, neither particle is detected, and so the
experimenter is unaware that a decay has taken place. In four of these
outcomes one of the particles is not detected. If the experimenter rejects
these data (in the belief that the apparatus is not functioning properly and
that if it had been functioning properly, the data recorded would have been
representative of the accepted data), he is left with the usual four
possible outcomes.'' Similarly, P. Grangier describes the meaning of fair
sampling assumption as [8]: ``The detection-efficiency loophole argues that,
in most experiments, only a very small fraction of the particles generated
are actually detected....So, to extract a meaningful conclusion from the
observed data, it was necessary to assume that a small fraction of data
provides a fair statistical sample''. Considering FS assumption, however, 
\textit{there is no reason why the data recorded are representative of the
accepted data, what is the nature of recording probabilities and how one can
interpret the efficiencies}. These questions are \textit{irrelevant} when
one refers to the FS assumption. In contrast, if we negate such a
hypothesis, i.e. if we believe that the rejected data may have a significant
role in calculating the correlations, it would be a crucial task to
elucidate the above points. This is our main concern in the following. (See
the appendix for a more concrete discussion about the FS assumption.)

Let us consider an actual double-channel Bell experiment where for each
emitted photon a binary event (i.e., passage or non-passage corresponding to
two polarization eigenvalues) occurs at each analyzer. We assume a
stochastic local hidden-variable (SLHV) theory, in which $\lambda $
represents a collection of hidden variables, belonging to a space $\Lambda $%
. To have a complete physical description of the whole system, the hidden
variables $\lambda $ are assumed to include the underlying variables of both
the particles and devices. At this level, $p_{r}^{(1)}(\widehat{a},\lambda )$
is the probability that the result $r$ is detected for the polarization of
the first photon along $\widehat{a}$, where $r=\pm 1$ corresponds to two
eigenvalues of photon's polarization and the angle $\widehat{a}$ is an angle
from the $x$-axis to the transmission axis of the first photon's
polarization filter. The detection probability $p_{r}^{(1)}(\widehat{a}%
,\lambda )$ can be defined as the following:

\begin{equation}
p_{r}^{(1)}(\widehat{a},\lambda )=p_{r,id}^{(1)}(\widehat{a},\lambda )\ \eta
_{1r}(\widehat{a},\lambda )  \tag{1}
\end{equation}
Here, $p_{r,id}^{(1)}(\widehat{a},\lambda )$ is the probability that if the
first photon encounters a polarizer at angle $\widehat{a}$, it will then be
detected in channel $r$ in an \textit{ideal} experiment. In an analogous
real experiment, we assume that $\eta _{1r}(\widehat{a},\lambda )$ denotes
the overall efficiency of detecting the first photon with polarization along 
$\widehat{a}$ in channel $r$. It contains, e.g., the probability that the
first photon reaches its detector and then will be detected with a definite
chance. One can define $p_{q}^{(2)}(\widehat{b},\lambda )$ in a similar
fashion for the second photon with $q=\pm 1$.

The probability of non-detection of photons \textsl{1} and \textsl{2}, along 
$\widehat{a}$ and $\widehat{b}$ respectively, are represented by $%
p_{0}^{(1)}(\widehat{a},\lambda )$ and $p_{0}^{(2)}(\widehat{b},\lambda )$,
where the index zero denotes non-detection. According to relation (1), $%
p_{0}^{(1)}(\widehat{a},\lambda )$ can be defined as $p_{0}^{(1)}(\widehat{a}%
,\lambda )=1-\alpha (\widehat{a},\lambda )$, where $\alpha (\widehat{a}%
,\lambda )=\stackunder{r=\pm 1}{\sum }\ p_{r}^{(1)}(\widehat{a},\lambda )$
is a representative function of the overall hidden efficiency$\ \eta _{1r}(%
\widehat{a},\lambda )$. If one assumes that $\eta _{1r}(\widehat{a},\lambda
) $ is independent of the measured value of the polarization $r$ (which
means that all the efficiencies are the same at two channels $+$ and $-$),
then $\alpha (\widehat{a},\lambda )=\ \eta _{1}(\widehat{a},\lambda )$ can
be interpreted as an overall measure of the efficiency at the
hidden-variable level. I.e.,

\begin{equation}
p_{0}^{(1)}(\widehat{a},\lambda )=1-\eta _{1}(\widehat{a},\lambda )  \tag{2}
\end{equation}
A similar relation can be considered for $p_{0}^{(2)}(\widehat{b},\lambda )$.

The joint probability for detection of the two photons with the outcomes $r$
and $q$ corresponding to the polarizations along $\widehat{a}$ and $\widehat{%
b}$, respectively, is assumed to be:

\begin{equation}
p_{rq}^{(12)}(\widehat{a},\widehat{b},\lambda )=p_{r}^{(1)}(\widehat{a}%
,\lambda )\ p_{q}^{(2)}(\widehat{b},\lambda )  \tag{3}
\end{equation}

This is known as Bell's locality condition [10]. Relations similar to (3)
hold for the joint probabilities concerning non-detections.

In a SLHV theory, the average value of the outcomes of polarizations of two
photons along $\widehat{a}$ and $\widehat{b}$ is given by

\begin{eqnarray}
\varepsilon ^{(12)}(\widehat{a},\widehat{b},\lambda ) &=&\stackunder{r,q=\pm
1}{\sum }rq\ p_{rq}^{(12)}(\widehat{a},\widehat{b},\lambda )  \nonumber \\
&=&\varepsilon ^{(1)}(\widehat{a},\lambda )\ \varepsilon ^{(2)}(\widehat{b}%
,\lambda )  \tag{4}
\end{eqnarray}
where $\varepsilon ^{(1)}(\widehat{a},\lambda )$ and $\varepsilon ^{(2)}(%
\widehat{b},\lambda )$ are the average values of the outcomes of
polarizations for photons \textsl{1} and \textsl{2} along $\widehat{a}$ and $%
\widehat{b}$, respectively. Assuming that the above probabilities are
normalized to one, we have:

\begin{equation}
\stackunder{j=\pm 1,0}{\sum }\ p_{j}^{(1)}(\widehat{a},\lambda )=\stackunder{%
j=\pm 1,0}{\sum }\ p_{j}^{(2)}(\widehat{b},\lambda )=1  \tag{5}
\end{equation}
Now, it is obvious that

\begin{equation}
0\leq p_{r}^{(1)}(\widehat{a},\lambda )\leq \alpha (\widehat{a},\lambda ) 
\tag{6}
\end{equation}
and

\begin{equation}
0\leq p_{q}^{(2)}(\widehat{b},\lambda )\leq \beta (\widehat{b},\lambda ) 
\tag{7}
\end{equation}
where $\alpha (\widehat{a},\lambda )=\stackunder{r=\pm 1}{\sum }\
p_{r}^{(1)}(\widehat{a},\lambda )$ and $\beta (\widehat{b},\lambda )=%
\stackunder{q=\pm 1}{\sum }\ p_{q}^{(2)}(\widehat{b},\lambda )$. The
constraints (6) and (7) are actual constraints for the detection of single
particles at the hidden-variable level. In the ideal limit, where $\alpha (%
\widehat{a},\lambda )\rightarrow 1$ and $\beta (\widehat{b},\lambda
)\rightarrow 1$, the probability of detection lies within the interval $%
\left[ 0,1\right] $. Using the aforementioned constraints, one gets:

\begin{equation}
\mid \varepsilon ^{(1)}(\widehat{a},\lambda )\mid \leq \alpha (\widehat{a}%
,\lambda )  \tag{8}
\end{equation}
and

\begin{equation}
\mid \varepsilon ^{(2)}(\widehat{b},\lambda )\mid \leq \beta (\widehat{b}%
,\lambda )  \tag{9}
\end{equation}

In the following, we introduce three independent solutions which include
some plausible assumptions about the nature of \textit{non-detection}
probabilities at the level of hidden variables as well as the relation of
the empirical correlations with the predictions of a SLHV theory. These
assumptions provide alternative ways for deriving an \textit{efficient} type
of CHSH inequality. Then, one can argue for the soundness of the recent
photonic experiments. Yet, there are some important points which should be
noted here. Our solutions I and II below involve assumptions about the
probabilities of non-detection. The non-detection probabilities are
unobservable and it has been usually recommended to avoid them. Thus, the
earlier works in this area involved constraints about the probabilities of
detection, rather than non-detection [7, 10]. Nevertheless, it is reasonable
to think that a more plausible approach with \textit{weaker} assumptions is
achieved when one takes into account the non-detection events. This is the
main point of the present work in which the nature of the auxiliary
assumptions are completely different with the so-called fair sampling or
no-enhancement assumptions in CHSH or CH inequalities.

As indicated before, what we are proposing here is that the non-detection
probabilities \textit{do} have an important role in calculating the photonic
correlations. But, we shall argue that there are situations in which one can
define an \textit{effective} correlation function only based on \textit{%
detected} events and derive an inequality which only contains the so-called
effective correlations. This is our purposed inequality. Here, we survey
these situations in the context of the following solutions.

\textbf{Solution I. }This is based on the assumption that \textit{at the
level of hidden variables, the probability of non-detection of each
individual photon is independent of the direction of its polarization filter}%
, i.e.,

\begin{equation}
p_{0}^{(1)}(\widehat{a},\lambda )=p_{0}^{(1)}(\widehat{a^{\prime }},\lambda
);  \tag{10-a}
\end{equation}

\begin{equation}
p_{0}^{(2)}(\widehat{b},\lambda )=p_{0}^{(2)}(\widehat{b^{\prime }},\lambda )
\tag{10-b}
\end{equation}
According to relation (1), this means also that for each individual photon,
the hidden probabilities for reaching a detector and detecting by it should
be independent of the earlier preparation made by choosing a definite
polarization angle.

Now, let us consider the set of polarization directions $\widehat{a},$ $%
\widehat{a^{\prime }}$ for the first photon and $\widehat{b},$ $\widehat{%
b^{\prime }}$ for the second one. Furthermore, we define the function $u$ as

\begin{equation}
u:=x(y-y^{\prime })+x^{\prime }(y+y^{\prime })  \tag{11}
\end{equation}
where $x:=\varepsilon ^{(1)}(\widehat{a},\lambda ),$ $x^{\prime
}:=\varepsilon ^{(1)}(\widehat{a^{\prime }},\lambda ),$ $y:=\varepsilon
^{(2)}(\widehat{b},\lambda )$ and $y^{\prime }:=\varepsilon ^{(2)}(\widehat{%
b^{\prime }},\lambda )$. We have also $\mid x\mid \leq \alpha ,$ $\mid
x^{\prime }\mid \leq \alpha ^{\prime },$ $\mid y\mid \leq \beta $ and $\mid
y^{\prime }\mid \leq \beta ^{\prime }$, in which for example $\alpha
:=\alpha (\widehat{a},\lambda )=1-p_{0}^{(1)}(\widehat{a},\lambda )$, $\beta
:=\beta (\widehat{b},\lambda )=1-p_{0}^{(2)}(\widehat{b},\lambda )$ and
similar definitions hold for $\alpha ^{\prime }$ and $\beta ^{\prime }.$
Considering the relations (10-a) and (10-b), we have $\alpha =\alpha
^{\prime }=\alpha (\lambda )$ and $\beta =\beta ^{\prime }=\beta (\lambda )$%
. So, the limits of $\mid x\mid $ and $\mid x^{\prime }\mid $ as well as $%
\mid y\mid $ and $\mid y^{\prime }\mid $ are the same. Since $u$ is a linear
function of the variables $x,$ $x^{\prime },$ $y$ and $y^{\prime },$ its
upper and lower bounds are determined by the limits of these variables. The
bounds are tabulated in the Table 1. This table shows that $u$ is confined
by the limits $2\alpha \beta $ and $-2\alpha \beta $. \bigskip

\begin{tabular}{|c|c|c|c|c|c|}
\hline
$Rows$ & $x$ & $x^{\prime }$ & $y$ & $y^{\prime }$ & $u$ \\ \hline
1 & $-\alpha $ & $-\alpha ^{\prime }$ & $-\beta $ & $-\beta ^{\prime }$ & $%
\alpha (\beta -\beta ^{\prime })+\alpha ^{\prime }(\beta +\beta ^{\prime
})=2\alpha \beta $ \\ \hline
2 & $\alpha $ & $-\alpha ^{\prime }$ & $-\beta $ & $-\beta ^{\prime }$ & $%
-\alpha (\beta -\beta ^{\prime })+\alpha ^{\prime }(\beta +\beta ^{\prime
})=2\alpha \beta $ \\ \hline
3 & $-\alpha $ & $\alpha ^{\prime }$ & $-\beta $ & $-\beta ^{\prime }$ & $%
\alpha (\beta -\beta ^{\prime })-\alpha ^{\prime }(\beta +\beta ^{\prime
})=-2\alpha \beta $ \\ \hline
4 & $-\alpha $ & $-\alpha ^{\prime }$ & $\beta $ & $-\beta ^{\prime }$ & $%
-\alpha (\beta +\beta ^{\prime })-\alpha ^{\prime }(\beta -\beta ^{\prime
})=-2\alpha \beta $ \\ \hline
5 & $-\alpha $ & $-\alpha ^{\prime }$ & $-\beta $ & $\beta ^{\prime }$ & $%
\alpha (\beta +\beta ^{\prime })+\alpha ^{\prime }(\beta -\beta ^{\prime
})=2\alpha \beta $ \\ \hline
6 & $\alpha $ & $\alpha ^{\prime }$ & $-\beta $ & $-\beta ^{\prime }$ & $%
-\alpha (\beta -\beta ^{\prime })-\alpha ^{\prime }(\beta +\beta ^{\prime
})=-2\alpha \beta $ \\ \hline
7 & $\alpha $ & $-\alpha ^{\prime }$ & $\beta $ & $-\beta ^{\prime }$ & $%
\alpha (\beta +\beta ^{\prime })-\alpha ^{\prime }(\beta -\beta ^{\prime
})=2\alpha \beta $ \\ \hline
8 & $\alpha $ & $-\alpha ^{\prime }$ & $-\beta $ & $\beta ^{\prime }$ & $%
-\alpha (\beta +\beta ^{\prime })+\alpha ^{\prime }(\beta -\beta ^{\prime
})=-2\alpha \beta $ \\ \hline
9 & $-\alpha $ & $\alpha ^{\prime }$ & $\beta $ & $-\beta ^{\prime }$ & $%
-\alpha (\beta +\beta ^{\prime })+\alpha ^{\prime }(\beta -\beta ^{\prime
})=-2\alpha \beta $ \\ \hline
10 & $-\alpha $ & $\alpha ^{\prime }$ & $-\beta $ & $\beta ^{\prime }$ & $%
\alpha (\beta +\beta ^{\prime })-\alpha ^{\prime }(\beta -\beta ^{\prime
})=2\alpha \beta $ \\ \hline
11 & $-\alpha $ & $-\alpha ^{\prime }$ & $\beta $ & $\beta ^{\prime }$ & $%
-\alpha (\beta -\beta ^{\prime })-\alpha ^{\prime }(\beta +\beta ^{\prime
})=-2\alpha \beta $ \\ \hline
12 & $\alpha $ & $\alpha ^{\prime }$ & $\beta $ & $-\beta ^{\prime }$ & $%
\alpha (\beta +\beta ^{\prime })+\alpha ^{\prime }(\beta -\beta ^{\prime
})=2\alpha \beta $ \\ \hline
13 & $\alpha $ & $\alpha ^{\prime }$ & $-\beta $ & $\beta ^{\prime }$ & $%
-\alpha (\beta +\beta ^{\prime })-\alpha ^{\prime }(\beta -\beta ^{\prime
})=-2\alpha \beta $ \\ \hline
14 & $\alpha $ & $-\alpha ^{\prime }$ & $\beta $ & $\beta ^{\prime }$ & $%
\alpha (\beta -\beta ^{\prime })-\alpha ^{\prime }(\beta +\beta ^{\prime
})=-2\alpha \beta $ \\ \hline
15 & $-\alpha $ & $\alpha ^{\prime }$ & $\beta $ & $\beta ^{\prime }$ & $%
-\alpha (\beta -\beta ^{\prime })+\alpha ^{\prime }(\beta +\beta ^{\prime
})=2\alpha \beta $ \\ \hline
16 & $\alpha $ & $\alpha ^{\prime }$ & $\beta $ & $\beta ^{\prime }$ & $%
\alpha (\beta -\beta ^{\prime })+\alpha ^{\prime }(\beta +\beta ^{\prime
})=2\alpha \beta $ \\ \hline
\end{tabular}
\bigskip

\textsl{Table 1: The limits of }$\QTR{sl}{u}$\textsl{.\bigskip }

Thus, under these conditions, we have:

\begin{equation}
\mid u\mid \leq 2\alpha \beta  \tag{12}
\end{equation}
In the ideal limit we have $\mid u\mid \leq 2$.

Now, we assume that the empirical correlation functions have a definite
relation with the averages of the outcomes of polarizations of the two
photons along certain directions in a SLHV theory. For example, for the two
polarization directions $\widehat{a}$ and $\widehat{b}$, we define:

\begin{equation}
E^{(12)}(\widehat{a},\widehat{b})=\int_{\Lambda }\varepsilon ^{(1)}(\widehat{%
a},\lambda )\ \varepsilon ^{(2)}(\widehat{b},\lambda )\ \rho (\lambda )\
d\lambda  \tag{13}
\end{equation}
where, $E^{(12)}(\widehat{a},\widehat{b})$ is the correlation function of
the polarization measurements of the two photons along $\widehat{a}$ and $%
\widehat{b},$ and $\rho (\lambda )$ is the normalized probability density of 
$\lambda $ over $\Lambda .$ Using the definitions of $\alpha $ and $\beta ,$
we have:

\begin{equation}
\stackunder{r,q=\pm 1}{\sum }P_{rq}^{(12)}=\int_{\Lambda }\alpha (\lambda )\
\beta (\lambda )\ \rho (\lambda )\ d\lambda  \tag{14}
\end{equation}
where $P_{rq}^{(12)}$ is the probability of the simultaneous detection of
the outcome $r$ for the first photon and $q$ for the second photon, with
polarizations along two arbitrary directions, in a real experiment. The
relation (14) is independent of the polarization directions. But, this does
not mean that the total number of photons recorded by each detector is
independent of the directions of the polarization filters, because the
number of undetected photons has a weighty role in the definition of the
detection probabilities.

Using the relations (11), (13) and (14), the inequality (12) takes the
following form:

\begin{equation}
\mid U\mid \leq M  \tag{15}
\end{equation}
where

\begin{equation}
U=E^{(12)}(\widehat{a},\widehat{b})-E^{(12)}(\widehat{a},\widehat{b^{\prime }%
})+E^{(12)}(\widehat{a^{\prime }},\widehat{b})+E^{(12)}(\widehat{a^{\prime }}%
,\widehat{b^{\prime }})  \tag{16}
\end{equation}
and $M=2\stackunder{r,q=\pm 1}{\sum }P_{rq}^{(12)}$.

In an ideal case, we have $M=2$ where by ideal we mean an experiment in
which the probabilities of non-detection are zero. In view of the fact that
in general $M\leq 2$, one can infer from (15) that

\begin{equation}
\mid U\mid \leq 2  \tag{17}
\end{equation}
The inequality (17) is known as CHSH inequality in the literature.

Now, we define the effective correlation functions measured in the photonic
experiments as

\begin{equation}
E_{eff}^{(12)}(\widehat{a},\widehat{b}):=\frac{E^{(12)}(\widehat{a},\widehat{%
b})}{\stackunder{r,q=\pm 1}{\sum }P_{rq}^{(12)}(\widehat{a},\widehat{b})}=%
\frac{\stackunder{r,q=\pm 1}{\sum }rq\ N_{rq}^{(12)}(\widehat{a},\widehat{b})%
}{\stackunder{r,q=\pm 1}{\sum }N_{rq}^{(12)}(\widehat{a},\widehat{b})} 
\tag{18}
\end{equation}
where $N_{rq}^{(12)}(\widehat{a},\widehat{b})$ is the number of photons that
are detected with the outcomes $r$ and $q$ along $\widehat{a}$ and $\widehat{%
b}$, respectively. Assuming that $P_{00}^{(12)}(\widehat{a},\widehat{b}%
)=P_{0}^{(1)}(\widehat{a})P_{0}^{(2)}(\widehat{b})$, we have$\stackunder{%
r,q=\pm 1}{\sum }P_{rq}^{(12)}(\widehat{a},\widehat{b})=\left( 1-P_{0}^{(1)}(%
\widehat{a})\right) \left( 1-P_{0}^{(2)}(\widehat{b})\right) ,$where $%
P_{0}^{(1)}(\widehat{a})$ ($P_{0}^{(2)}(\widehat{b})$) is the probability of
non-detection of photon \textsl{1 }(\textsl{2}) with a polarization along $%
\widehat{a}$ ($\widehat{b}$) and $P_{00}^{(12)}(\widehat{a},\widehat{b})$ is
the joint probability of non-detection for both photons. Using the
definition (18), the inequality (15) is reduced to

\begin{equation}
\mid U_{eff}\mid \leq 2  \tag{19}
\end{equation}
where

\begin{equation}
U_{eff}=E_{eff}^{(12)}(\widehat{a},\widehat{b})-E_{eff}^{(12)}(\widehat{a},%
\widehat{b^{\prime }})+E_{eff}^{(12)}(\widehat{a^{\prime }},\widehat{b}%
)+E_{eff}^{(12)}(\widehat{a^{\prime }},\widehat{b^{\prime }})  \tag{20}
\end{equation}

The inequality (19) is our proposed Bell-type inequality for a real
experiment. This is the inequality which has been tested in recent photonic
experiments and is shown to have been violated.

\textbf{Solution II. }The first solution was based on assumptions that are
used at the hidden-variable level. In the second solution, however, both the
experimental and hidden-variable levels are under consideration. To derive
(19), we assume that \textit{non-detection probabilities for each individual
photon are the same at both levels}, i.e.,

\begin{equation}
P_{0}^{(1)}(\widehat{a})=p_{0}^{(1)}(\widehat{a},\lambda );\quad P_{0}^{(1)}(%
\widehat{a^{\prime }})=p_{0}^{(1)}(\widehat{a^{\prime }},\lambda ) 
\tag{21-a}
\end{equation}

\begin{equation}
P_{0}^{(2)}(\widehat{b})=p_{0}^{(2)}(\widehat{b},\lambda );\quad P_{0}^{(2)}(%
\widehat{b^{\prime }})=p_{0}^{(2)}(\widehat{b^{\prime }},\lambda ) 
\tag{21-b}
\end{equation}

Here, one can argue that non-detection probabilities are hidden, as is the
case at the hidden-variable level. Because, there is no way for their
detection. The necessary condition for the acceptance of above relations is
the assumption that the non-detection probability for each individual
photon, at the hidden-variable level, is independent of $\lambda $\textit{. }%
Or, equivalently, this means that the hidden efficiencies for reaching a
detector and detection by it are the same as the experimental ones (see
relation (1)).

Subsequently, One can define an effective average value at the level of
hidden variables, as

\begin{eqnarray}
\varepsilon _{eff}^{(12)}(\widehat{a},\widehat{b},\lambda ) &=&\stackunder{%
r,q=\pm 1}{\sum }rq\ (\frac{p_{r}^{(1)}(\widehat{a},\lambda )}{1-p_{0}^{(1)}(%
\widehat{a},\lambda )})(\frac{p_{q}^{(2)}(\widehat{b},\lambda )}{%
1-p_{0}^{(2)}(\widehat{b},\lambda )})  \nonumber \\
&=&\varepsilon _{eff}^{(1)}(\widehat{a},\lambda )\ \varepsilon _{eff}^{(2)}(%
\widehat{b},\lambda )  \tag{22}
\end{eqnarray}
where $\varepsilon _{eff}^{(1)}(\widehat{a},\lambda )=\stackunder{r=\pm 1}{%
\sum }r(\frac{p_{r}^{(1)}(\widehat{a},\lambda )}{1-p_{0}^{(1)}(\widehat{a}%
,\lambda )})$ and $\varepsilon _{eff}^{(2)}(\widehat{b},\lambda )=%
\stackunder{q=\pm 1}{\sum }q(\frac{p_{q}^{(2)}(\widehat{b},\lambda )}{%
1-p_{0}^{(2)}(\widehat{b},\lambda )})$. Using (6) and (7), we get:

\begin{equation}
\left| \varepsilon _{eff}^{(1)}(\widehat{a},\lambda )\right| \leq 1  \tag{23}
\end{equation}

\begin{equation}
\left| \varepsilon _{eff}^{(2)}(\widehat{b},\lambda )\right| \leq 1  \tag{24}
\end{equation}

Using relations (22)-(24) and integrating over $\lambda $, one can prove
(19), in a fashion similar to the proof of CHSH inequality. Based on the
relations (21-a) and (21-b), the function $E_{eff}^{(12)}(\widehat{a},%
\widehat{b})$ in (18) has the following relation with the hidden variables
level:

\begin{equation}
E_{eff}^{(12)}(\widehat{a},\widehat{b})=(\frac{1}{1-P_{0}^{(1)}(\widehat{a})}%
)(\frac{1}{1-P_{0}^{(2)}(\widehat{b})})\int_{\Lambda }\varepsilon ^{(1)}(%
\widehat{a},\lambda )\ \varepsilon ^{(2)}(\widehat{b},\lambda )\ \rho
(\lambda )\ d\lambda  \tag{25}
\end{equation}

\textbf{Solution III. }Unlike the first and second solutions, here, we do
not make any assumption about the nature of the non-detection probabilities.
Instead, we make a conjecture that one can replace (13) by

\begin{equation}
E_{eff}^{(12)}(\widehat{a},\widehat{b})=\int_{\Lambda }\varepsilon
_{eff}^{(1)}(\widehat{a},\lambda )\ \varepsilon _{eff}^{(2)}(\widehat{b}%
,\lambda )\ \rho (\lambda )\ d\lambda  \tag{26}
\end{equation}
where $\varepsilon _{eff}^{(1)}(\widehat{a},\lambda )$ and $\varepsilon
_{eff}^{(2)}(\widehat{b},\lambda )$ are defined as before and $%
E_{eff}^{(12)}(\widehat{a},\widehat{b})$ is defined in as (18).

One can prove the inequality (19) by using (22)-(24) and (26). The relations
(13) and (26) are identical in the ideal limit, but they have different
predictions for the real experiments. The physical content of the relation
(26) is that \textit{one can always reproduce experimental results using the
predictions of a SLHV theory}, whereas relations like (13) indicate that in
real experiments one cannot reproduce the predictions of quantum mechanics
without making extra assumptions.

Our three solutions for reproducing the inequality (19) involve compatible
assumptions. The conjunction of the first two solutions means that the
probability of non-detection for a given particle should be merely a
function of instrumental efficiencies. Then, the relations (25) and (26) are
obtained by dividing both sides of (13) by a detection constant. In such a
case, it is assumed that non-detection is only an instrumental problem which
is present but does not depend on what a microphysical theory is aimed to
describe.

\section{The Predictions of Quantum Mechanics}

What are the predictions of quantum mechanics for the inequality (19)? In a
real double-channel experiment, the respective quantum mechanical joint
probability for detecting two photons is nearly equal to [15]:

\begin{equation}
P_{rq,QM}^{(12)}(\widehat{a},\widehat{b})\approx \frac{1}{4}\eta _{1}\eta
_{2}f_{12}\left[ 1+rq\ F\cos 2(\widehat{a}-\widehat{b})\right]  \tag{27}
\end{equation}
In this relation, $\eta _{k}$ is the efficiency of detecting the $k$th
photon ($k=1,2$). The function $f_{12}=f_{1}f_{2}$ shows the probability
that both photons reach their detectors, where $f_{1}$ denotes the
probability for the first photon reaching its corresponding detector and $%
f_{2}$ is the same probability for the second photon. They are indicating
the efficiencies of the two corresponding collimators for photons \textsl{1}
and \textsl{2}. The function $F$ is a measure of the correlation of the two
emitted photons. In the relation (27), the efficiencies of the analyzers are
assumed to be approximately perfect, which is the case in all recent
photonic experiments. In an ideal experiment, all of the above efficiencies
are equal to one. Here, for simplicity, we assume that $\eta _{1}\approx
\eta _{2}\approx \eta $. Then, using (27), we obtain:

\begin{equation}
\stackunder{r,q=\pm 1}{\sum }P_{rq,QM}^{(12)}(\widehat{a},\widehat{b}%
)\approx \eta ^{2}f_{12}  \tag{28}
\end{equation}
which is independent of polarization directions. Since $\left(
1-P_{0,QM}^{(1)}\right) \approx \eta f_{1}$ and $\left(
1-P_{0,QM}^{(2)}\right) \approx \eta f_{2},$ the relation (28) is also equal
to $\left( 1-P_{0,QM}^{(1)}\right) \left( 1-P_{0,QM}^{(2)}\right) $.

Now, using the fact that $E_{QM}^{(12)}(\widehat{a},\widehat{b})=\stackunder{%
r,q=\pm 1}{\sum }rq\ P_{rq,QM}^{(12)}(\widehat{a},\widehat{b}),$ the quantum
correlation function for the polarization directions $\widehat{a}$ and $%
\widehat{b}$ is: 
\begin{equation}
E_{QM}^{(12)}(\widehat{a},\widehat{b})\approx \eta ^{2}f_{12}F\cos 2(%
\widehat{a}-\widehat{b})  \tag{29}
\end{equation}
and subsequently,

\begin{equation}
E_{QM,eff}^{(12)}(\widehat{a},\widehat{b})=\dfrac{E_{QM}^{(12)}(\widehat{a},%
\widehat{b})}{\stackunder{r,q=\pm 1}{\sum }P_{rq,QM}^{(12)}(\widehat{a},%
\widehat{b})}\approx F\cos 2(\widehat{a}-\widehat{b})  \tag{30}
\end{equation}

If we choose $\mid \widehat{a}-\widehat{b}\mid =$ $\mid \widehat{a^{\prime }}%
-\widehat{b}\mid =$ $\mid \widehat{a^{\prime }}-\widehat{b^{\prime }}\mid
=\varphi $ and $\mid \widehat{a}-\widehat{b^{\prime }}\mid =3\varphi $, then
(19) yields

\begin{equation}
F\mid 3\cos \varphi -\cos 3\varphi \mid \leq 2  \tag{31}
\end{equation}
For $\varphi =\frac{\pi }{4},$ we have $F\sqrt{2}\leq 1$. In real
experiments where the entangled photon pairs are produced through
spontaneous parametric down-conversion, $F$ is about $0.95$ or more [16,
17]. Since the inequality (31) is independent of the efficiency of detectors
and collimators (two main facts responsible for the FS assumption), the
predictions of quantum mechanics violate (19) and thus (31), without using
the FS assumption. This may be the reason why in spite of the low
efficiencies in Bell's photonic experiments, the value of $U_{eff}$ in (19)
agrees so well with predictions of the standard quantum mechanics and why
this value is nearly the same in different experiments with different
efficiency factors.\bigskip

\textbf{Appendix\bigskip }

Here, we want to elucidate the meaning of FS assumption more clearly. The
CHSH inequality can be expressed as

\begin{equation}
\mid U\mid \leq 2  \tag{A-1}
\end{equation}
where $U$ is a linear combination of some empirical correlation functions
along different directions defined in (16). The main issue of FS assumption
is that one can use the inequality (19) $\mid U_{eff}\mid \leq 2$ instead of
(A-1) where $U_{eff}$ is defined in (20).

But, how is it possible to obtain (19) from (A-1)\textit{\ }and what is the
role of FS assumption in deriving (19)? To answer these two questions, we
first remember that an effective correlation function measured in the
photonic experiments can be defined as (18). Now, it is obvious that for
every $\widehat{k}=\widehat{a}$ or $\widehat{a^{\prime }}$ and $\widehat{l}=%
\widehat{b}$ or $\widehat{b^{\prime }}$, we should have $\left|
E_{eff}^{(12)}(\widehat{k},\widehat{l})\right| \geq \left| E^{(12)}(\widehat{%
k},\widehat{l})\right| $, because $\stackunder{r,q=\pm 1}{\sum }%
P_{rq}^{(12)}(\widehat{k},\widehat{l})\leq 1$. We reformulate $E^{(12)}(%
\widehat{k},\widehat{l})$ as $E^{(12)}(\widehat{k},\widehat{l}%
)=E_{eff}^{(12)}(\widehat{k},\widehat{l})-\epsilon _{kl}$, where $\epsilon
_{kl}:=\frac{E^{(12)}(\widehat{k},\widehat{l})\left( 1-\stackunder{r,q=\pm 1%
}{\sum }P_{rq}^{(12)}(\widehat{k},\widehat{l})\right) }{\stackunder{r,q=\pm 1%
}{\sum }P_{rq}^{(12)}(\widehat{k},\widehat{l})}$. Consequently, one can
begin with the CHSH inequality (A-1) and reach the following one

\begin{equation}
-2+\epsilon \leq U_{eff}\leq 2+\epsilon  \tag{A-2}
\end{equation}
where $\epsilon =\epsilon _{ab}-\epsilon _{a^{\prime }b}+\epsilon
_{ab^{\prime }}+\epsilon _{a^{\prime }b^{\prime }}$. Since, there is no way
to measure $E^{(12)}(\widehat{k},\widehat{l})$ in real photonic experiments,
the empirical value of $\epsilon $ cannot be determined. Considering the
predictions of quantum mechanics (see section 3), however, one can show that 
$\epsilon _{QM}=(1-\eta ^{2}f_{12})U_{eff}$, where $\eta $ and $f_{12}$ are
some efficiencies defined in section 3. Thus, what quantum mechanics
predicts is that $\left| U_{eff,QM}\right| \leq \dfrac{2}{\eta ^{2}f_{12}}$
which is far from violation in actuality.

Yet, there are two situations in which $\epsilon $ can be assumed to be zero:

\begin{quote}
1- The experiment is performed under \textit{ideal} conditions, that is $%
\stackunder{r,q=\pm 1}{\sum }P_{rq}^{(12)}$ reaches one actually.

2- The statistics of the experiment can be \textit{fairly} constructed on
the basis of the accessible data , that is all the predicted values will
remain valid when $\stackunder{r,q=\pm 1}{\sum }P_{rq}^{(12)}$ is
renormalized to unity (FS assumption).
\end{quote}

There is no way to obtain (19) from (A-1) except for the above conditions.
No assumption about the nature of \textit{non}-detection probabilities does
help. The renormalization of $\stackunder{r,q=\pm 1}{\sum }P_{rq}^{(12)}$ as
well as the validness of the predictions under new conditions are the key
points.

Nevertheless, it is logically possible to derive (19) by a \textit{different}
approach. One possible way is to begin with an alternative inequality $\mid
U\mid \leq 2\stackunder{r,q=\pm 1}{\sum }P_{rq}^{(12)}$ (relation (15))
which is the basis of our first solution. In our two other solutions, we
suppose some alternative relations for $E_{eff}^{(12)}(\widehat{a},\widehat{b%
})$ corresponding to the hidden-variable level, to derive (19). So, our
three suggested solutions impose more stringent conditions on a SLHV theory
for being compatible with the experiments.

\end{document}